\documentstyle[prl,aps,twocolumn,graphics,epsfig]{revtex}

\begin{document}
\title{High-efficiency quantum interrogation measurements via the quantum
Zeno effect}
\author{P. G. Kwiat$^{1,*}$, 
A. G. White$^{1}$, 
J. R. Mitchell$^{1}$, O. Nairz$^{2,\dagger}$, G. Weihs$^{2,\dagger}$, 
H. Weinfurter$^{2,\ddagger}$, and A. Zeilinger$^{2,\dagger}$}
\address{$^{1}$Physics Division, P-23, Los Alamos National Laboratory,
Los Alamos, New Mexico 87545}

\address{$^{2}$Institute for Experimental Physics, 
University of Innsbruck, Innsbruck 6020, Austria}

\date{to appear in Phys. Rev. Lett; submitted June 11, 1999}

\draft
\maketitle
%\vskip -0.1cm
\begin{abstract}
The phenomenon of quantum interrogation allows one to optically detect 
the presence of an absorbing object, without the measuring light 
interacting with it.  In an application of the quantum Zeno effect, 
the object inhibits the otherwise coherent evolution of the light, 
such that the probability that an interrogating photon is absorbed can 
in principle be arbitrarily small.  We have implemented this 
technique, demonstrating efficiencies exceeding the 50\% 
theoretical-maximum of the original ``interaction-free'' measurement 
proposal.
%Here we present a specific embodiment based on the quantum Zeno effect, 
%specifically, an inhibited polarization rotation. This permitted the
%first observation of interaction-free measurement ``efficiency'' $>50\%$, 
%the upper threshold of the original proposal.  
We have also predicted and experimentally verified a previously 
unsuspected dependence on loss; efficiencies of up to 73\% were 
observed and the feasibility of efficiencies up to 85\% was 
demonstrated.

\end{abstract}
\draft
%\vspace {1 cm}
\pacs{PACS numbers: 03.65.Bz,03.65.-a,42.50.-p,42.25.Hz}
\vspace{-0.4 cm}

``Negative result'' measurements were discussed by 
Renninger~\cite{Renninger} and later by Dicke~\cite{Dicke}, who 
analyzed the change in an atom's wavefunction by the {\it 
nonscattering} of a photon from it.  In 1993 Elitzur and Vaidman (EV) 
showed that the wave-particle duality of light could allow 
``interaction-free'' quantum interrogation of classical objects, in 
which the presence of a non-transmitting object is ascertained 
seemingly without interacting with it~\cite{EV1}, i.e., with no photon 
absorbed or scattered by the object.  In the basic EV technique, an 
interferometer is aligned to give complete destructive interference in 
one output port -- the ``dark'' output -- in the absence of an object.  
The presence of an opaque object in one arm of the interferometer 
eliminates the possibility of interference so that a photon may now be 
detected in this output.  If the object is completely 
non-transmitting, any photon detected in the dark output port must 
have come from the path {\it not} containing the object.  Hence, the 
measurements were deemed ``interaction-free'', though we stress that 
this term is sensible only for objects that completely block the beam.  
For measurements on partially-transmitting (and quantum) objects, we 
suggest the more general terminology ``quantum interrogation''.  In 
any event there is necessarily a coupling between light and object 
(formally describable by some interaction Hamiltonian) -- somewhat 
paradoxically, in the high-efficiency schemes discussed below, it is 
crucial that the {\it possibility} of an interaction exist, in order 
to reduce the probability that such an interaction actually occurs.

The EV gedanken experiment has been realized using true single-photon 
states~\cite{KwiatPRL} and with a classical light beam attenuated to 
the single-photon level~\cite{duMarchie}, as well as in neutron 
interferometry~\cite{Hafner}.  This methodology has even been employed 
to investigate the possibility of performing ``absorption-free'' 
imaging~\cite{IFMPRA}.  The EV technique suffers two serious 
drawbacks, however.  First, the measurement result is ambiguous at 
least half of the time -- a photon may be detected in the non-dark 
output port whether or not there is an object.  Second, at most half 
of the measurements are interaction-free~\cite{KwiatPRL,IFMPRA}.  
Following Elitzur and Vaidman~\cite{EV1}, we define a figure of merit 
$\eta = \rm{P(QI)}/[ \rm{P(QI)}+ \rm{P(abs)}]$ to characterize the 
``efficiency'' of a given scheme, where P(QI) is the probability that 
the photon is detected in the otherwise dark port, and P(abs) is 
the probability that the object absorbs or scatters the photon.  
Physically, $\eta$ is the {\it fraction} of measurements that are 
``interaction-free''.  The maximum achievable efficiency, obtained by 
adjusting the reflectivities of the EV interferometer beamsplitters, 
is $\eta = 50\%$~\cite{EV1,KwiatPRL,IFMPRA}.

It was proposed that one could circumvent these limitations by using a 
hybrid scheme~\cite{KwiatPRL}, combining the interferometric ideas of 
EV and incorporating an optical version of the {\it quantum Zeno 
effect}~\cite{Misra}, in which a weak, repeated measurement inhibits 
the otherwise coherent evolution of the interrogating photon.
%Here we present a 
%specific embodiment based on an inhibited polarization rotation.
Our specific embodiment of the Zeno effect is based on an inhibited 
polarization rotation~\cite{Peres}, although the only generic 
requirement is a weakly-coupled multi-level system.  A photon with 
\begin{figure}[ht!]
\begin{center}
\epsfxsize=0.6 \columnwidth
\epsfbox{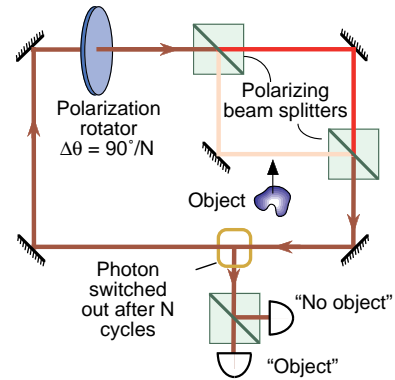}
\end{center}
\footnotesize
\caption{Simple schematic of a hybrid scheme to allow high-efficiency 
quantum interrogation of the presence of an opaque object.  With no 
object, the initial horizontal polarization of the interrogating 
photon is rotated stepwise to vertical.  The presence of an object in 
the {\bf V}-arm inhibits this evolution via the optical quantum Zeno 
effect~\protect\cite{Peres}, so that the final polarization after $N$ 
cycles unambiguously indicates the presence or absence of the object: 
{\bf V} polarization $\rightarrow$ ``no object''; {\bf H} polarization 
$\rightarrow$ ``object''.}
\end{figure}
\noindent horizontal ({\bf H}) polarization is directed through a 
series of $N$ polarization rotators (e.g., optically active elements), 
each of which rotates the polarization by $\Delta \theta \equiv 
\pi/2N$.  The net effect of the entire stepwise quantum evolution is 
to rotate the photon's polarization to vertical ({\bf V}).  We may 
inhibit this evolution if at each stage we make a measurement of the 
polarization in the {\bf H}/{\bf V} basis, e.g., by inserting a 
horizontal polarizer after each rotator.  Since the probability of 
being transmitted through each polarizer is just $\cos^2\Delta 
\theta$, the probability P(QI) of being transmitted through all $N$ of 
them is simply $\cos ^{2N}(\Delta \theta) \approx 1 - \pi^2/4N$, and 
the complementary probability of absorption is P(abs) $\approx 
\pi^2/4N$.  Thus, increasing the number of cycles leads to an 
arbitrarily small probability that the photon is ever absorbed.

Obviously the Zeno phenomenon as described is of limited use, because 
it requires polarizing objects.  Figure 1 shows the basic concept to 
allow quantum interrogation of {\it any} non-transmitting object.  A 
single photon is made to circulate $N$ times through the setup, before 
it is somehow removed and its polarization analyzed.  As in the 
example above, the photon, initially {\bf H}-polarized, is rotated by 
$\Delta \theta = \pi/2N$ on each cycle, so that after $N$ cycles the 
photon is found to have {\bf V} polarization.  This rotation is 
unaffected by the polarization-interferometer (consisting of two 
polarizing beam splitters, which ideally transmit all {\bf 
H}-polarized and reflect all {\bf V}-polarized light; and two 
identical-length arms), which simply separates the light into its {\bf 
H} and {\bf V} components and adds them back with the same relative 
phase.  If there is an object in the vertical arm of the 
interferometer, however, only the {\bf H} component of the light is 
passed; i.e., each {\it non}-absorption by the object [with 
probability $\cos ^2\Delta \theta$] projects the wavefunction back 
into its initial state.  Hence, after $N$ cycles, either the photon 
will still have {\bf H} polarization [with probability P(QI)], 
unambiguously indicating the presence of the object, or the object 
will have absorbed the photon [probability P(abs)].  By going to 
higher $N$, P(abs) can in principle be made arbitrarily small.  In the 
absence of any losses or other non-idealities, $\eta = $P(QI), so 
that $\eta \rightarrow 1$ as $N \rightarrow \infty$.

Demonstrating this phenomenon in an actual experiment required several 
modifications (see Fig.  2).  A horizontally-polarized laser pulse was 
coupled into the system by a highly reflective mirror.  The light was 
attenuated so that the average photon number per pulse after the 
mirror was between 0.1 and 0.3.  The photon then bounced between this 
recycling mirror and one of the mirrors making up a polarization 
Michelson interferometer.  At each cycle a waveplate rotated the 
polarization by $\Delta \theta$.  After the desired number of cycles 
$N$, the photon was switched out of the system by applying a 
high-voltage pulse to a Pockels cell in each interferometer arm, 
thereby rotating the polarization of the photon by $90^{\circ}$, so 
that it exited via the other port of the polarizing beam splitter.  
The exiting photon was then analyzed by 
\begin{figure}[ht!]
\begin{center}
\epsfxsize=0.7 \columnwidth
\epsfbox{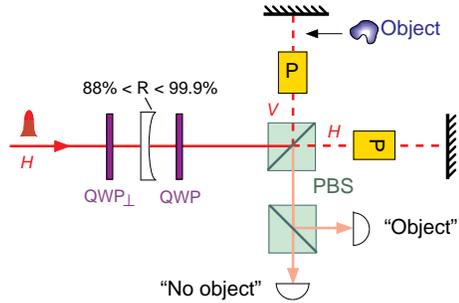}
\end{center}
\footnotesize
\caption{Experimental system to demonstrate high-efficiency quantum 
interrogation.  Photons from a pulsed laser at 670nm are coupled into 
the recycling system via a high-reflectivity recycling mirror 
(initially flat, later curved; see Fig.  3).  A double pass through 
the quarter waveplate (QWP) served to rotate the polarization by a 
fixed amount each cycle; an extra waveplate (QWP$_{\bot}$) in the 
entrance beam was used to compensate for the initial pass.  On each 
cycle the photon passed through a polarization interferometer (with a 
polarizing beamsplitter [PBS]); to fine tune the interferometer phase, 
one mirror was mounted on a piezoelectric ``bimorph''.  The Pockels 
cells (P) were used to switch the photons out after a desired number of 
cycles -- a $\sim 3$ kV pulse was applied, which after the double pass 
rotated the polarization of the photon by $90^{\circ}$, so that it 
exited via the other port of the PBS. The exiting photon was then 
analyzed by the adjustable polarizer and single-photon detector (EG\&G 
\#SPCM-AQ-141, preceded by an interference filter [10nm FWHM, centered 
at 670nm] to reduce background).  The final polarization of the 
detected photon indicates the presence ({\bf V}-polarized) or absence 
({\bf H}-polarized) of an object in the reflected arm of the 
interferometer.  (Not shown: active feedback Helium Neon laser which 
ran below the plane of the 670nm light, to stabilize the 
interferometer.)}
\end{figure}
\noindent an adjustable polarizer and single-photon detector.  With no 
object, the polarization was found to be essentially horizontal, 
indicating that the stepwise rotation of polarization had taken place 
(remember, the final polarization is $90^{\circ}$ rotated by the 
Pockels cell).  With the object in the vertical-polarization arm of 
the interferometer, this evolution was inhibited, and a photon exiting 
the system was vertically-polarized, an interaction-free measurement 
of the presence of the object~\cite{practical_footnote}.

A number of intermediate configurations were investigated before 
arriving at the arrangement described above~\cite{KwiatNobel}.  With 
these the {\it feasibility} of quantum interrogation with $\eta$ up to 
85\% was inferred (for a hypothetically lossless system) -- there was 
no way to directly measure the amount of light absorbed by the object.  
In the present experiment, we made a {\it direct} measurement of the 
probability that a photon took the object path, by applying a constant 
voltage to the Pockels cell in that path, thereby directing these 
photons to the single-photon detector at each cycle.  With the DC 
voltage applied, photons exiting with {\bf H} polarization correspond 
to P(abs), while those with {\bf V} polarization (which exit only 
after $N$ cycles) correspond to P(QI).  (We verified that the 
\begin{figure}[ht!]
\begin{center}
\epsfxsize=0.7 \columnwidth
\epsfbox{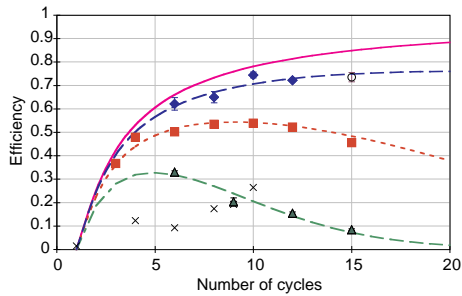}
\end{center}
\footnotesize
\caption{Efficiency versus number of cycles $N$ for several system 
configurations.  The curves are theoretical predictions, based on the 
measured losses for each configuration.  The triangles and the 
dot-dashed curve correspond to a lossy non-switching system in which 
the photons experienced 8\% loss/cycle due to the input coupler, and 
leaked out through a flat 88\%-reflective output coupler; the squares 
and dotted curve correspond to the system in Fig.  2, with a somewhat 
lossy Pockels cell in the no-object arm ($T$ = 95.1\%), and a flat 
recycling mirror ($R$ = 96.2\%); the diamonds and the dashed curve 
correspond to a better Pockels cell ($T$ = 97.7\%), and a curved 
recycling mirror ($R$ = 97.4\%); and the circle corresponds to a 
higher reflectivity ($R$ = 99.4\%) curved mirror.  The solid curve is 
the prediction for a lossless system.  Several representative 
measurements of the ``noise'' in our quantum interrogation process are 
also shown (crosses).}
\end{figure}
\noindent rates corresponding to P(QI) were similar whether using the 
DC-biased Pockels cell as the object, or physically blocking that arm 
of the interferometer.)  Rather unexpectedly, the efficiencies 
determined in this fashion were significantly lower than both the 
theoretical predictions and the previous inferred measurements, which 
agreed well with each other.  The reason is that the effects of {\it 
loss} in the system were normalized out in the previous 
measurements~\cite{KwiatNobel}.

That loss should reduce the actual efficiencies was somewhat 
surprising, since losing a photon from the system seems equivalent to 
never sending it in initially.  This line of reasoning is faulty: A 
photon contributing to P(QI) must necessarily remain in the system 
for all $N$ cycles, thus sampling any loss $N$ times; in contrast, a 
photon contributing to P(abs) could be absorbed on any cycle, hence 
remains in the system on average less than $N$ cycles, and sees less 
loss than a photon contributing to P(QI).  The net effect is that, 
whereas $\eta \rightarrow 1$ for a large number of lossless cycles, in 
the presence of loss $\eta$ reaches a maximum value less than one 
before falling again toward zero~\cite{etalossy}.  This places a 
strong constraint on the achievable efficiencies in any real system.

Figure 3 shows the experimental verification of this phenomenon, as 
well as the modified theoretical predictions, which are in good 
agreement.  Despite the efficiency reduction due to loss, we were able 
to observe $\eta$'s of up to $74.4 \pm 1.2\%$.  Also shown in Fig.  3 
are several representative measurements of the ``noise'' of our 
quantum interrogation system, from events in which an object was 
indicated (i.e., photons were detected with vertical polarization) 
even though none was actually present.  The main causes were 
imperfections of the optical elements, and interferometer instability, 
despite active stabilization.

Because the same photon detector was used to determine both P(QI) and 
P(abs) in our measurements, the detector efficiency factors out of the 
calculation for $\eta$.  When our highest-observed value of $\eta$ is 
corrected for our finite detection 
efficiencies~\cite{trueetafootnote}, we arrive at an adjusted $\eta$ 
of $53.1 \pm 1.6\%$, where we have included the effects of both the 
detector efficiency (65\%) and the 10-nm filter (60\% transmission) 
used to reduce background.  Because this value of $\eta$ is only 
marginally above the 50\% threshold of the original EV scheme, we also 
took one set of data in which the 10-nm filter was removed.  Our 
measured $\eta$ was $72.3 \pm 1.1\%$, implying a raw efficiency of 
{\bf $62.9 \pm 1.3\%$}; that is, in measurements with the opaque 
object, $\sim 2/3$ of the photons performed an ``interaction-free'' 
measurement, and $\sim 1/3$ did not~\cite{practical_footnote}; in 
other words, the object's presence can be unambiguously determined 
while absorbing only ``1/3 of a photon''.  This is, to our knowledge, 
one of the first practical utilizations of the quantum Zeno effect.

A wholly different method of quantum interrogation, relying on 
disrupting the resonance condition of a high-finesse cavity (and hence 
called ``resonance interaction-free measurement"), has been 
proposed~\cite{HPaul} and recently demonstrated~\cite{Tsegaye}, with 
efficiencies similar to those reported here.  If the cavity mirrors 
both have reflectivity $R$, a narrow bandwidth photon incident on the 
empty cavity can have a near-unity probability of transmission, i.e., 
a detector observing the reflection from the entrance mirror to the 
cavity will never fire, in principle.  An object in the cavity will 
naturally prevent the resonance condition (this can be thought of as 
an impedance mismatch), so the detector in the reflected mode will 
detect the photon with probability P(QI) = $R$, while the object will 
have a probability 1-$R$ of absorbing the photon.  The efficiency of 
this scheme ($R$, in the ideal case) can thus exceed the EV 50\% 
threshold, like the method based on the Zeno effect.  However, the two 
techniques have very different characteristics.  For example, while 
the Zeno technique employs broadband photon wavepackets, the resonance 
methods necessarily require a very narrow frequency spectrum for the 
interrogating photons.  Because of the pulsed nature of the Zeno 
effect, the duration of the experiment is precisely fixed (to $N$ 
cycles); the duration of the measurement with the cavity is less well 
defined, determined by the ring-down time of the cavity.  Conversely, 
it is relatively easy to allow photons to leak out of a cavity, 
whereas actively switching them as in our scheme is experimentally 
more challenging.  Finally, as presented here, both techniques require 
interferometric stability; however, this is not strictly necessary for 
the Zeno method if one has a polarizing object, e.g., a 
polarization-selective atomic transition.

Achieving higher efficiencies with these techniques will require 
increasing the working number of cycles $N$.  However, the performance 
of the system becomes increasingly more sensitive to optical 
imperfections and to interferometer instability.  The effect of loss 
is also multiplied.  We believe that with sufficient engineering these 
problems could be reduced, allowing operation at up to O(100) cycles 
or higher, giving efficiencies $>93\%$~\cite{whythis}.  Finally, {\it 
crosstalk} in the polarizing beamsplitter (i.e., not all horizontal 
polarized light is transmitted, and not all vertical polarized light 
is reflected, about $\sim 1\%$ in our present system) must be kept to 
a minimum.  In particular, we observed spurious interference effects 
when the reflection probability [$\sin^2(\pi/2N)$] becomes comparable 
to the
%PBS
crosstalk. Use of birefringent material polarizers, whose crosstalk 
figures are $\sim 10^{-5}$, may mitigate this problem.

If the efficiencies can be improved as discussed above, one can 
envision using the methods to examine quantum mechanical objects, such 
as an atom or ion, one of whose states couples to the interrogating 
light (``object''), and another of whose states does not couple (``no 
object'').  In the simplest situation we can determine which state the 
system is in with a greatly reduced probability of exciting it out of 
that state.  Such a process might be called ``absorption-free 
spectroscopy'', and could be useful for studying photosensitive 
systems.  More interestingly, when the quantum system is in a 
superposition of the two states, the light becomes {\it entangled} 
with the quantum system~\cite{EV1,Entangle,Karlsson,Geszti}.  Such an 
effect may have use as a quantum ``wire'', e.g., as an interface for 
connecting together two quantum computers~\cite{Zhen}.  Finally, in 
the limit as $\eta \rightarrow 1$, these techniques of quantum 
interrogation will function even if there are several photons (or an 
{\it average} of several photons, as in a weak coherent state).  It 
should then be possible to produce Schr\"{o}dinger-cat like states 
$\alpha |VVV \ldots \rangle + \beta |HHH \ldots \rangle$, where $ |VVV 
\ldots\rangle$ ($ |HHH \ldots \rangle$) represents several photons 
with vertical (horizontal) polarization~\cite{Entangle}.  Such states 
would have great interest for studying the classical-quantum boundary, 
and the phenomenon of decoherence.

We would like to acknowledge I. Appelbaum, N. Kurnit, G. Peterson, V. 
Sandberg, and C. Simmons for help with our feedback system and Pockels 
cells; ON, GW, HW, and AZ acknowledge support by the Austrian Science 
Foundation FWF project number S6502.  $^*$ kwiat@lanl.gov.  $^\dagger$ 
Current address: Institute for Experimental Physics, University of 
Vienna, Wien 1090, Austria.  $^{\ddagger}$ Current address: Sektion 
Physik, Ludwig-Maximilians-Universit\"{a}t, M\"{u}nchen Schellingstr.  
4/III D-80799 M\"{u}nchen.

%\vspace{2.25 in}
%

%\vspace{3 in}
%

%

%\vspace{2.5 in}


\vspace{-0.5 cm}
\begin{references}
\vspace{-1.8 cm}    

\bibitem{Renninger} M. Renninger, Z. Phys.  {\bf 158}, 417 (1960).

\bibitem{Dicke} R. H. Dicke, Am.  J. Phys.  {\bf 49}, 925 (1981).

\bibitem{EV1} A. Elitzur and L. Vaidman, Found.  Phys.  {\bf 23}, 987 
(1993).

\bibitem{KwiatPRL} P. G. Kwiat, H. Weinfurter, T. Herzog, A. 
Zeilinger, and M. A. Kasevich, Phys.  Rev.  Lett.  {\bf 74}, 4763 
(1995).

\bibitem{duMarchie} E. H. du Marchie van Voorthuysen, Am.  J. Phys.  
{\bf 64}, 1504 (1996).

\bibitem{Hafner} M. Hafner and J. Summhammer, Phys.  Lett.  A {\bf 
235}, 563 (1997).

\bibitem{IFMPRA} A. G. White, J. R. Mitchell, O. Nairz, and P. G. 
Kwiat, Phys.  Rev.  A {\bf 58}, 605 (1998).

\bibitem{Misra} B. Misra and E. C. G. Sudarshan, J. Math Phys.  {\bf 
18}, 756 (1977).

\bibitem{Peres} A. Peres, Am.  J. Phys.  {\bf 48}, 931 (1980).

\bibitem{practical_footnote} In a true practical implementation, one 
would stop the input pulses as soon as a photon was detected,
% at one of the output ports, 
thereby preventing the possibility that the object might absorb a 
photon from a subsequent pulse.  While this was not done for our 
proof-of-principle experiment, there is no reason (aside from 
interferometer drift issues) that the system should behave differently 
if the interval between pulses were, say, 100s (time to 
position/remove the suspect object), instead of the 0.001s we used.  
But it would then have taken weeks to accumulate sufficient statistics 
to accurately determine $\eta$.

\bibitem{KwiatNobel} P. G. Kwiat, Phys.  Scripta {\bf T76}, 115 
(1998).

\bibitem{etalossy} As before, $\eta = \rm{P(QI)}/[ \rm{P(QI)}+ 
\rm{P(abs)}]$, where now
\begin{eqnarray}
\rm{P(QI)} & = &
\left( T_{empty} \rm{cos}^2 \Delta \theta \right)^{N}
\left( T_{rec.} \right)^{N-1}  \nonumber \\ 
\rm{P(abs)} & = & \frac {T_{obj} {\rm sin}^2 \Delta \theta
[1 - (T_{empty} T_{rec.}\rm{cos}^{2} \Delta \theta)^{N}]}
{(1 - T_{empty}T_{rec.} \rm{cos}^{2} \Delta \theta)}  \nonumber \;;
%\label{??}
\end{eqnarray}
$N$ is the number of cycles, $\Delta \theta = \pi/2N$, and 
$T_{empty}$, $T_{obj}$, and $T_{rec.}$ are respectively the 
single-cycle transmission probabilities for the empty interferometer 
arm, the arm with the object, and the recycling arm ($\equiv T_{QWP}^2 
R_{mirror}$, with $R_{mirror}$ the recycling mirror reflectivity).  
See also J.-S. Jang, Phys.  Rev.  A {\bf 59}, 2322 (1999).

\bibitem{trueetafootnote} If the observed efficiency is $\eta_{obs}$, 
and the net detection efficiency is $\epsilon$, then the adjusted 
$\eta$ is given by \mbox{$\eta_{obs} \epsilon/(1 - \eta_{obs} (1 - 
\epsilon))$}.

\bibitem{HPaul} H. Paul and M. Pavicic, Int.  J. Theor.  Phys.  {\bf 
35}, 2085 (1996).

\bibitem{Tsegaye} T. Tsegaye, {\it et al.}, Phys.  Rev.  A {\bf 57}, 
3987 (1998).

\bibitem{whythis} This estimate assumes a single-pass Pockel-cell 
transmission of 99.5\%, recycling mirror reflectivity of 99.9\%, and 
reflection losses of 0.05\% on the QWP and PBS surfaces (all 
obtainable specifications with current technology).  Modern 
stabilization techniques for high-finesse optical cavities routinely 
achieve $>1000$ roundtrips.  The resulting maximum intrinsic efficiency 
of 94.5\% (at N = 107), is reduced to 93.2\% for detector efficiencies 
of 80\%~\cite{trueetafootnote}.

\bibitem{Entangle} P. G. Kwiat, H. Weinfurter, and A. Zeilinger, in 
{\it Coherence and Quantum Optics VII}, ed.  by J. Eberly, L. Mandel, 
and E. Wolf (Plenum Publishing Corp.  1996), p.  673.

\bibitem{Karlsson} A. Karlsson, G. Bjork, and E. Forsberg, Phys.  Rev.  
Lett.  {\bf 80}, 1198 (1998); G. Krenn, J. Summhammer, and K. Svozil, 
Phys.  Rev.  A {\bf 53}1228 (1998).

\bibitem{Geszti} T. Geszti, Phys. Rev. A {\bf 58}, 4206 (1998).

\bibitem{Zhen} J. Zhen and G. Guang-Can, Acta Phys.  Sinica {\bf 7}, 
437 (1998); A. G. White, P. G. Kwiat, and D. F. V. James, in {\it 
Proceedings of the Workshop on Mysteries, Puzzles, and Paradoxes in 
Quantum Mechanics}, ed.  by R. Bonifacio (American Institute of 
Physics, New York, 1999), p.  268.

\end{references}
\end{document}